\documentclass[10pt,onecolumn,twoside]{article}

\usepackage{epsfig}
\usepackage{titlesec}
\usepackage{url} 
\usepackage{booktabs} 
\usepackage{amsfonts} 
\usepackage{amssymb} 
\usepackage{nicefrac} 
\usepackage{microtype} 
\usepackage{mathrsfs}
\usepackage{bm} 
\usepackage{cite} 
\usepackage{comment}
\usepackage{graphicx} 
\usepackage{mathtools} 
\usepackage{algpseudocode}
\usepackage{algorithm}
\usepackage{amsthm} 
\usepackage{enumitem} 
\usepackage{multirow} 
\usepackage{tabularx} 
\usepackage{hhline}
\usepackage{arydshln} 
\usepackage{enumitem} 
\usepackage{siunitx}
\usepackage[font=small,skip=5pt]{caption}
\usepackage{parskip}
\usepackage{comment}
\usepackage[dvipsnames]{xcolor}
\usepackage{fullpage}
\usepackage[hidelinks]{hyperref}

\algtext*{EndIf}
\algtext*{EndFor}
\algtext*{EndWhile}

\newcolumntype{L}[1]{>{\raggedright\arraybackslash}p{#1}}
\newcolumntype{C}[1]{>{\centering\arraybackslash}p{#1}}
\newcolumntype{R}[1]{>{\raggedleft\arraybackslash}p{#1}}

\theoremstyle{plain} 

\def\defn{\,\coloneqq\,}
\def\argmin{\mathop{\mathsf{arg\,min}}} 

\def\lim{\mathop{\mathsf{lim}}} 

\def\prox{\mathsf{prox}}

\def\zer{\mathsf{zer}}
\def\fix{\mathsf{fix}}

\def\ebm{{\bm{e}}}

\def\sbm{{\bm{s}}}
\def\xbm{{\bm{x}}}

\def\ybm{{\bm{y}}}
\def\zbm{{\bm{z}}}

\def\nbm{{\bm{n}}}

\def\thetabm{{\bm{\theta}}}
\def\zerobm{\bm{0}}

\def\Hbm{{\bm{H}}}

\def\Fbm{{\bm{F}}}
\def\Sbm{{\bm{S}}}
\def\Pbm{{\bm{P}}}

\def\xbmast{{\bm{x}^\ast}}

\def\xbmhat{{\widehat{\bm{x}}}}

\def\xbfhat{{\widehat{\mathbf{x}}}}
\def\xbfhat{{\widehat{\bm{x}}}}

\def\Tsf{{\mathsf{T}}}

\def\Dsf{{\mathsf{D}}}

\def\Gsf{{\mathsf{G}}}

\def\Rsf{{\mathsf{R}}}

\def\C{\mathbb{C}}

\def\N{\mathbb{N}}

\def\Lcal{{\mathcal{L}}}

\definecolor{pink}{HTML}{000000}
\definecolor{lightgreen}{RGB}{229,251,229}
\definecolor{red}{rgb}{0,0,0}

\title{RARE: Image Reconstruction using Deep Priors Learned\\without Groundtruth}

\author{Jiaming~Liu%
\thanks{Department of Electrical \& Systems Engineering, Washington University in St.~Louis, St.~Louis, MO 63130.}
\hspace{0.05em},
Yu~Sun%
\thanks{Department of Computer Science \& Engineering, Washington University in St.~Louis, St.~Louis, MO 63130.}
\hspace{0.05em}, 
Cihat~Eldeniz%
\thanks{Mallinckrodt Institute of Radiology, Washington University School of Medicine, St.~Louis, MO 63110.}
\hspace{0.05em},
Weijie~Gan$^{\dagger}$,\\
Hongyu~An$^{\ddagger,}$%
\thanks{Department of Biomedical Engineering and the Division of Biology and Biomedical Sciences, Washington University in St.~Louis, St.~Louis, MO 63130, USA.}
, and Ulugbek~S.~Kamilov$^{\ast, \dagger}$}

\begin{document}
\date{}
\maketitle

\begin{abstract}
\emph{Regularization by denoising (RED)} is an image reconstruction framework that uses an image denoiser as a prior. Recent work has shown the state-of-the-art performance of RED with \emph{learned denoisers} corresponding to pre-trained convolutional neural nets (CNNs). In this work, we propose to broaden the current denoiser-centric view of RED by considering priors corresponding to networks trained for more general artifact-removal. The key benefit of the proposed family of algorithms, called \emph{regularization by artifact-removal (RARE)}, is that it can leverage priors learned on datasets containing only undersampled measurements. This makes RARE applicable to problems where it is practically impossible to have fully-sampled groundtruth data for training. We validate RARE on both simulated and experimentally collected data by reconstructing a free-breathing whole-body 3D MRIs into ten respiratory phases from heavily undersampled k-space measurements. Our results corroborate the potential of learning regularizers for iterative inversion directly on undersampled and noisy measurements.
\end{abstract}

\section{Introduction}

Image reconstruction from a set of undersampled observations is central for various imaging applications. The reconstruction is often formulated as an optimization problem
\begin{equation}
\label{Eq:RegularizedOptimization}
\xbmhat = \argmin_{\xbm} f(\xbm) \quad\text{with}\quad f(\xbm) = g(\xbm) + h(\xbm),
\end{equation}
where $\xbm$ denotes the unknown image, $g$ is the data-fidelity term that relies on the measurement model to impose data-consistency, and $h$ is the imaging prior used to regularize the solution for ill-posed problems. Over the years, many regularizers have been proposed as imaging priors, including those based on transform-domain sparsity, low-rank penalty, and dictionary learning~\cite{Rudin.etal1992, Figueiredo.Nowak2001, Figueiredo.Nowak2003, Hu.etal2012, Elad.Aharon2006, Degraux.etal2017, Kamilov2017}.

There has been considerable interest in leveraging image denoisers as imaging priors within iterative reconstruction. \emph{Regularization by denoising (RED)}~\cite{Romano.etal2017} is a recent framework that uses an off-the-shelf denoiser to specify an explicit regularizer that has a simple gradient. It has been shown that RED leads to excellent performance in various imaging problems, when equipped with advanced denoisers~\cite{Romano.etal2017, Reehorst.Schniter2019, Mataev.etal2019}. Similar results have been observed in a related class of algorithms known as \emph{plug-and-play priors (PnP)}~\cite{Venkatakrishnan.etal2013, Sun.etal2019a, Meinhardt.etal2017, Ryu.etal2019, Song.etal2020, Tirer.Giryes2019}. Recent work has additionally shown the effectiveness of RED for imaging priors specified as pre-trained denoising \emph{convolutional neural nets (CNNs)}~\cite{Sun.etal2019c, Metzler.etal2018, Wu.etal2019}. It has been generally observed that learned denoisers are essential for achieving the state-of-the-art results in many contexts. The training of corresponding priors, however, can be a significant practical challenge in some applications, where the measurement setup fundamentally limits the ability to collect fully-sampled groundtruth data.

In this work, we broaden the current denoiser-centric view of RED by proposing the \emph{regularization by artifact removal (RARE)} framework for iterative image reconstruction. RARE extends RED beyond priors trained for the removal of additive white Gaussian noise (AWGN) by considering those trained for more general artifact removal. The ability of RARE to leverage any artifact-removal CNN within iterative reconstruction naturally leads to a new paradigm for learning imaging priors that requires no fully-sampled groundtruth data. Inspired by Noise2Noise~\cite{Lehtinen.etal2018}, we propose a scheme, called \emph{Artifact2Artifact (A2A)}, for training an imaging prior by mapping pairs of artifact and noise contaminated images obtained directly from undersampled measurements. We demonstrate the practical relevance of RARE by reconstructing 4D images in free-breathing motion-compensated MRI, where it is practically impossible to acquire fully-sampled training data. Our results on both simulated and experimental data corroborate the ability of RARE to effectively leverage priors learned on datasets containing only undersampled and noisy measurements. RARE thus addresses an important gap in the current literature on RED by providing a flexible, scalable, and theoretically sound algorithm applicable to a wide variety of imaging problems.

The outline for the rest of the paper is as follows. In Section~\ref{Sec:Background}, we introduce our notation and review some relevant background material on image reconstruction, discussing both the traditional model-based and the recent learning-based approaches. In Section~\ref{Sec:Proposed}, we present and explain the details of RARE, including our strategy for training the prior without groundtruth data. In Section~\ref{Sec:Experiments}, we provide numerical experiments that corroborate the excellent performance and practical relevance of RARE. Section~\ref{Sec:Conclusion} concludes the paper.

\section{Background}
\label{Sec:Background}
\subsection{Imaging Inverse Problems}
Consider a linear inverse problem of recovering an unknown image $\xbm \in \C^n$ from its noisy measurements $\ybm \in \C^m$ specified by the linear system
\begin{equation}
\label{Eq:ForwardModel}
\ybm = \Hbm\xbm + \ebm
\end{equation}
where the measurement operator $\Hbm \in \C^{m \times n}$ characterizes the response of the imaging system, and $\ebm \in \C^m$ is the AWGN. For example, in \emph{compressed sensing (CS)} MRI~\cite{Lustig.etal2007, Lustig.etal2008} the goal is to recover an image $\xbm$ from sparsely-sampled Fourier measurements. In the special case of parallel MRI~\cite{Pruessmann.etal1999} with a dynamic object, the operator can be represented as
\begin{equation}
\Hbm_i^{(t)} =\Pbm^{(t)}\Fbm\Sbm_i, 
\end{equation}
where $\Fbm$ denotes the Fourier transform operator, $\Sbm_i$ is the matrix containing the per-pixel sensitivity map for the $i$th coil, and $\Pbm^{(t)}$ is the k-space sampling matrix at time $t$. If the k-space samples lie on the Cartesian grid, the fast Fourier transform (FFT) provides an efficient implementation of $\Fbm$ and its adjoint. If non-Cartesian k-space sampling is used, then a nonuniform FFT (NUFFT) is needed~\cite{Fessler.etal2003}. When the k-space is undersampled, the inverse problem is ill-posed even in the absence of noise; therefore, it is common to formulate the solution as a regularized inversion.

Conventional methods for reconstruction from underdetermined measurements rely on the sparsity of typical images in some transform domain, such as wavelet~\cite{Figueiredo.Nowak2003}, total variation (TV)~\cite{Rudin.etal1992}, or exploit the spatio-temporal correlations, such as in the k-t Sparse and Low-Rank (k-t SLR) model~\cite{lingalaetal.2011}.  Such regularizers generally result in non-smooth optimization problems. A variety of methods such as the proximal gradient method (PGM)~\cite{Kamilov.etal2017,Sun.etal2019b} and the alternating direction method of multipliers (ADMM)~\cite{Boyd.etal2011} have been developed for efficient minimization of nonsmooth functions, without differentiating them, by using the \emph{proximal operator}, defined as
\begin{equation}
\label{Eq:ProximalOperator}
\prox_{\mu h}(\zbm) \defn \argmin_{\xbm}\left\{\frac{1}{2}\|\xbm-\zbm\|_2^2 + \mu h(\xbm)\right\},
\end{equation}
where $\mu > 0$ is a parameter. Note that the proximal operator can be interpreted as the regularized image denoiser for AWGN with noise of variance of $\mu$.

\subsection{Deep Learning for Image Reconstruction}

Deep learning has recently gained popularity for solving imaging inverse problems. An extensive review of deep learning in the context of image reconstruction can be found in~\cite{McCann.etal2017, Lucas.etal2018, Knoll.etal2020}. Instead of explicitly defining an imaging prior, the traditional deep-learning approach is based on training an existing network architecture, such as UNet~\cite{Ronneberger.etal2015}, to invert the measurement operator by exploiting the natural redundancies in the imaging data~\cite{Mousavi.etal2015, Wang2016.etal, Kang.etal2017, DJin.etal2017, Han.etal2017, Chen.etal2017, Li.etal2018, Sun.etal2018, Lee.etal2018, Ye.etal2018, Yoo.etal2017}. It is common to first bring the measurements to the image domain and train the network to map the corresponding low-quality images $\{\xbfhat_j\}$ to their clean target versions $\{\xbm_j\}$ by solving an optimization problem
\begin{equation}
\label{Eq:cnn_loss}
\argmin_{\thetabm}\sum_{j}\Lcal (f_{\thetabm}(\xbfhat_{j}),\xbm_{j})),
\end{equation}
where $f_{\thetabm}$ represents the CNN parametrized by $\thetabm$, under the loss function $\Lcal$.  Popular loss functions include the $\ell_2$-norm and $\ell_1$-norm. In practice, \eqref{Eq:cnn_loss} can be optimized using the family of stochastic gradient descent (SGD) methods, such as adaptive moment estimation (ADAM)~\cite{Kingma.Ba2015}. For example, in the context of CS-MRI, previous methods have trained $f_\thetabm$ for mapping a zero-filled image $\xbmhat$ to a reconstruction $\xbm$ from a fully-sampled groundtruth data~\cite{Wang2016.etal, Han.etal2017}.

Noise2Noise (N2N)~\cite{Lehtinen.etal2018} is a recent methodology for addressing the problem of training deep neural networks on noisy labels.  In N2N, $f_\thetabm$ is trained to learn a mapping between pairs of images that have random and independent degradations. For example, given a family of noisy images $\{\xbm_j + \nbm_j, \xbm_j + \nbm^\prime_j\}$, with random independent vectors $\nbm_j$ and $\nbm'_j$, one can replace $(\xbm_j, \xbmhat_j)$ in~\eqref{Eq:cnn_loss} by the noisy pair and still learn the model $f_\thetabm$ that predicts a clean image. The key assumption for this strategy is to have the expected value of the noisy input to be equal to the clean signal~\cite{Lehtinen.etal2018}. This central idea has been recently expanded to other settings~\cite{Krull.etal2019, Batson.etal2019, Cha.etal2019}.

The idea of end-to-end learning can be refined to include the measurement operator in~\eqref{Eq:ForwardModel} (for an overview see \emph{``Unrolling''} in~\cite{McCann.etal2017} or \emph{``Neural networks and analytical methods''} in~\cite{Lucas.etal2018}). Inspired by LISTA~\cite{Gregor.LeCun2010}, the corresponding \emph{unfolding algorithms} interpret iterations of a regularized inversion as layers of a CNN and train it end-to-end in a supervised fashion~\cite{Schmidt.Roth2014, Chen.etal2015, Kamilov.Mansour2016, Bostan.etal2018}. In the context of CS-MRI, ADMM-Net~\cite{Yang.etal2016} and variational network (VN)~\cite{Hammernik.etal2018} have considered jointly training the image transforms and shrinkage functions within an unfolded algorithm. A related class of methods, such as CascadeNet~\cite{Schlemper.etal2018} and MoDL~\cite{Aggarwal.etal2019}, have considered including a full CNN as a trainable regularizer within an unfolded algorithm. Such unfolding algorithms have been shown to be effective in a number of problems~\cite{Biswas.etal2019, Hosseini.etal2019} and are closely related to the PnP/RED methods (discussed in Section~\ref{Sec:PnP}) that also combine the measurement operator and the imaging prior. However, the prior in PnP/RED is not trained end-to-end within an iterative algorithm, which significantly reduces the cost and memory requirements of training (see also the discussion in \emph{``Memory Requirements''} of~\cite{Schlemper.etal2018}). In particular, model-based end-to-end architectures require the storage of all the intermediary activation maps (which are 4D tensors for 3D CNNs) at every iteration of the unfolded algorithm, which leads to a significant memory burden for certain inverse problems where one needs to use higher dimensional or more complex CNNs.

\begin{figure*}[t]
\begin{center}
\includegraphics[width=16cm]{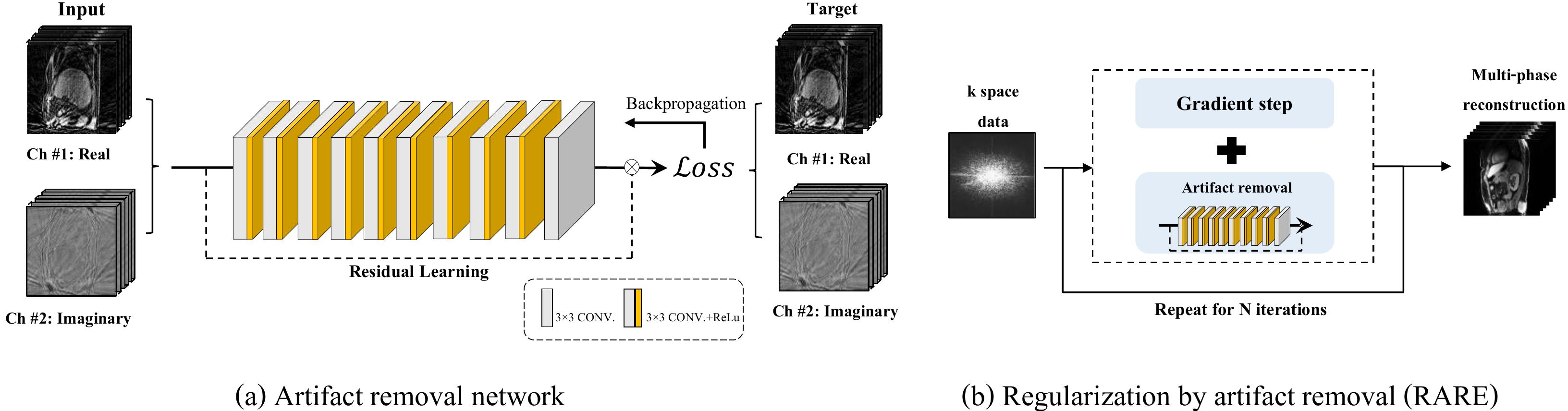}
\end{center}
\caption{Illustration of the proposed RARE framework on the problem of reconstructing 10 motion phases from a single free-breathing MRI acquisition. (a) The artifact-removal CNN is trained on complex-valued 3D images ($x$, $y$ for space and $p$ for phase) without any fully-sampled groundtruth. (b) RARE combines information from the CNN prior with that from the measurement operator to iteratively refine the solution.}
\label{Fig:pipline}
\end{figure*}

\subsection{Using Denoisers as Imaging Priors}
\label{Sec:PnP}

The mathematical equivalence between the proximal operator~\eqref{Eq:ProximalOperator} and image denoising has led to the development of denoiser-based iterative algorithms such as PnP~\cite{Venkatakrishnan.etal2013}  and RED~\cite{Romano.etal2017}. In PnP the proximal operator is replaced with an arbitrary image denoiser $\Dsf_\sigma$, where $\sigma > 0$ controls the strength of denoising. This simple replacement enables PnP to regularize the problem by using advanced denoisers, such as BM3D~\cite{Dabov.etal2007} and DnCNN~\cite{Zhang.etal2017}, that do not correspond to any explicit $h$. Recent studies have confirmed the state-of-the-art performance of PnP in many imaging applications. While the original formulation of PnP is based on ADMM, other algorithms have been developed using PGM and primal-dual splitting~\cite{Kamilov.etal2017, Ono2017, Meinhardt.etal2017}. 

RED is an alternative framework to PnP that uses an image denoiser to formulate an explicit regularization function
\begin{equation}
\label{Eq:Regularizer}
h(\xbm) = \frac{\tau}{2}\xbm^\Tsf(\xbm-\Dsf_\sigma(\xbm)),
\end{equation}
where $\tau > 0$ is the regularization parameter. When the denoiser is locally homogeneous and has a symmetric Jacobian~\cite{Romano.etal2017, Reehorst.Schniter2019}, the gradient of the RED regularizer $h$ has a simple form
\begin{align}
\label{Eq:FixedPoints}
\nabla h(\xbm) =  \tau (\xbm - \Dsf_\sigma(\xbm)),
\end{align}
which enables an efficient implementation of RED for problems of form~\eqref{Eq:RegularizedOptimization}. The excellent performance of RED under a learned denoiser has been demonstrated in super-resolution, phase retrieval, and compressed sensing~\cite{Metzler.etal2018, Wu.etal2019, Sun.etal2019c}. DeepRED~\cite{Mataev.etal2019} is a more recent extension of RED that relies on DIP~\cite{Ulyanov.etal2018} as an additional \emph{learning-free} CNN prior. The RARE framework introduced in the next section is a natural generalization of RED towards CNNs trained for artifact-removal, which enables it to leverage priors learned from undersampled and noisy data.

\section{Regularization by artifact removal}
\label{Sec:Proposed}

In this section, we describe the proposed framework, which represents a conceptual leap from denoising in RED to artifact removal. Remarkably, this conceptual leap is fully compatible with the key theoretical properties of the original RED framework, such as the existence of an explicit regularizer~\eqref{Eq:Regularizer} that has a simple gradient~\eqref{Eq:FixedPoints}, under the same assumptions described in~\cite{Romano.etal2017, Reehorst.Schniter2019}. RARE, however, significantly extends the practical applicability of RED by leveraging priors learned on datasets containing only undersampled and noisy measurements.

\subsection{Algorithmic Framework}

The CNN priors in the traditional formulation of RED are designed for AWGN removal. This is inconvenient for applications where it is difficult to directly synthesize the AWNG corrupted images for training the CNN. RARE (summarized in Algorithm~\ref{alg:rareagm}) circumvents this problem by replacing the denoiser $\Dsf_\sigma$ in RED by an artifact-removal network $\Rsf_\thetabm$, where $\thetabm$ denotes the weights of the CNN. RARE can thus leverage any existing image restoration CNN designed for removing artifacts specific to some imaging modality.

Let $\Hbm^\dagger$ denote the pseudoinverse of the measurement operator $\Hbm$. Then, given an initial solution provided by the artifact-removal network $\xbm^0 = \Rsf_\thetabm(\Hbm^\dagger\ybm)$, RARE iteratively refines it by infusing information from both the gradient of the data-fidelity term $\nabla g$ and the residual of the restoration CNN $(\xbm-\Rsf_\thetabm(\xbm))$. Specifically, RARE aims to find the solution $\xbmast$ within the zero set of the operator
\begin{align}
\label{Eq:ZeroSet}
&\zer(\Gsf) \defn \{\xbm \in \C^n: \Gsf(\xbm) = \zerobm\} \quad\text{with}\quad\\ &\quad\quad\Gsf(\xbm) \defn \nabla g(\xbm) + \tau(\xbm-\Rsf_\thetabm(\xbm)),\nonumber
\end{align}
where $\tau > 0$ is the regularization parameter.
Consider the following two sets
\begin{align*}
&\zer(\nabla g) \defn \{\xbm \in \C^n : \nabla g(\xbm) = \zerobm\}\quad\text{and}\\
&\fix(\Rsf_\thetabm) \defn \{\xbm \in \C^n : \xbm = \Rsf_\thetabm(\xbm)\},
\end{align*}
where $\zer(\nabla g)$ is the set of all critical points of the data-fidelity term and $\fix(\Rsf_\thetabm)$ is the set of all fixed points of the artifact-removal CNN. Intuitively, $\fix(\Rsf_\thetabm)$ correspond to all the vectors that are artifact-free according to the CNN, while $\zer(\nabla g)$ corresponds to the vectors that are consistent with the measured data. If $\xbmast \in \zer(\nabla g)\cap\fix(\Rsf_\thetabm)$, then $\Gsf(\xbmast) = \zerobm$ and $\xbmast$ is one of the solutions of RARE. Hence, any vector that is consistent with the data for a convex $g$ and artifact free according to $\Rsf_\thetabm$ is in the solution set of RARE. On the other hand, when $\zer(\nabla g)\cap \fix(\Rsf_\thetabm) = \varnothing$, then $\xbmast \in \zer(\Gsf)$ corresponds to an equilibrium point between these two sets, explicitly controlled via $\tau > 0$. Additionally, when $\Rsf_\thetabm$ satisfies the assumptions outlined in~\cite{Romano.etal2017, Reehorst.Schniter2019}, RARE also admits an explicit regularizer of form~\eqref{Eq:Regularizer} by replacing the denoiser $\Dsf_\sigma$ by the artifact-removal network $\Rsf_\thetabm$.

The simplest algorithm for computing~\eqref{Eq:ZeroSet} can be obtained by running the following fixed-point algorithm
\begin{align}
\label{Eq:RAREupdate}
\xbm^k &\leftarrow \xbm^{k-1} - \gamma \Gsf(\xbm^{k-1}),
\end{align}
where $\gamma>0$ is the step-size. RARE adds two notable extensions. First, we use the  backtracking line-search strategy~\cite{Boyd.Vandenberghe2004} to automatically adjust $\gamma > 0$. The line-search starts with some large step size, which is subsequently decreased by a factor $\beta \in (0,1)$ to ensure monotonic decrease $\|\Gsf(\xbm^k)\|_2 \leq \|\Gsf(\xbm^{k-1})\|_2$. Second, we equip RARE with Nesterov's acceleration~\cite{Nesterov2004} via $\{q_k\}$ for better convergence. The acceleration terms are updated in the usual way as
\begin{equation}
\label{Eq:qk}
q_k \leftarrow \frac{1}{2}(1 + \sqrt{1 + q_{k-1}^2}).
\end{equation}
Note that when $q_k = 1$ for all $k$, the algorithms reverts to the usual gradient method without acceleration. 

\begin{figure}[t]
\begin{center}
\begin{minipage}[t]{0.9\textwidth}
\begin{algorithm}[H]
\caption{$\mathsf{Regularization\;by\;Artifact\;Removal\; (RARE)}$}\label{alg:rareagm}
\begin{algorithmic}[1]
\State \textbf{input: } $\xbm^0 = \sbm^0$, $\gamma > 0$, $\tau > 0$, $\rho>0$, $\beta \in (0,1)$, and $\{q_k\}_{k \in \N}$
\For{$k = 1, 2, \dots$}
\State $\xbm^k \leftarrow \sbm^{k-1}-\gamma \Gsf(\sbm^{k-1})$\\
\quad\quad\quad \textsf{where} $\Gsf(\xbm) \defn \nabla g(\xbm) + \tau(\xbm-\Rsf_\thetabm(\xbm))$
\State $\sbm^k \leftarrow \xbm^k + ((q_{k-1}-1)/q_k)(\xbm^k-\xbm^{k-1}) $
\While{$\|\Gsf(\xbm^k)\|_2 > \|\Gsf(\sbm^{k-1})\|_2$}
\State $\gamma \leftarrow \beta \gamma$
\State $\xbm^k \leftarrow \sbm^{k-1}-\gamma \Gsf(\sbm^{k-1})$
\If{$\gamma< \rho$}
\State \textsf{break}
\EndIf
\EndWhile\label{euclidendwhile}
\EndFor\label{euclidendwhile}
\end{algorithmic}
\end{algorithm}
\end{minipage}
\end{center}
\end{figure}

\subsection{Learning without Groundtruth}

In many applications, it is practically impossible to obtain fully-sampled and noise-free acquisitions. However, it is generally straightforward to obtain several independent undersampled acquisitions of the same object. The \emph{Artifact2Artifact (A2A)} strategy described here enables one to train the imaging prior for RARE using only pairs of undersampled and noisy measurements. Consider a set of unknown images $\{\xbm_j\}$, each acquired multiple times
$$\ybm_{ij} = \Hbm_{ij} \xbm_j + \ebm_{ij},$$
where $i$ is the index of the specific acquisition and $\Hbm_{ij}$ denotes the specific measurement operator. One can then generate a dataset of artifact-contaminated images by simply applying the pseudoinverse of each measurement operator
\begin{equation}
\label{Eq:TrainData}
\xbmhat_{ij} = \Hbm_{ij}^\dagger\ybm_{ij}.
\end{equation}
The artifact-removal network can then be trained via the following empirical risk minimization
\begin{equation}
\label{Eq:N2Nlearn}
\argmin_{\thetabm}\sum_{i,j,i^\prime, j^\prime}\Lcal (\Rsf_{\thetabm}(\xbfhat_{ij}),\xbmhat_{i^\prime j^\prime})),
\end{equation}
where the goal is to simply map pairs of artifact-contaminated images in~\eqref{Eq:TrainData}. Note that, unlike the setting considered in Noise2Noise~\cite{Lehtinen.etal2018}, the artifacts in~\eqref{Eq:N2Nlearn} are not assumed to be AWGN, since each training label is assumed to be obtained from noisy sparsely-sampled measurements. However, the whole dataset (consisting of multiple acquisition of multiple objects) is assumed to compliment the information missing in each individual image, thus enabling the training of the imaging prior. Underlying assumption of A2A is that the expected value of the images $\{\xbmhat_{ij}\}_j$ with various acquisition times still matches the groundtruth vector $\xbm_i$. While this assumption might seem idealized for some practical applications, our experiments corroborate its excellent performance in the context of a heavily undersampled 4D MR image reconstruction under object motion. In particular, we show that the inclusion of the A2A trained $\Rsf_\thetabm$ within RARE updates that also include $\nabla g$ lead to a powerful image reconstruction strategy.

\begin{figure*}[t]
\begin{center}
\includegraphics[width=16cm]{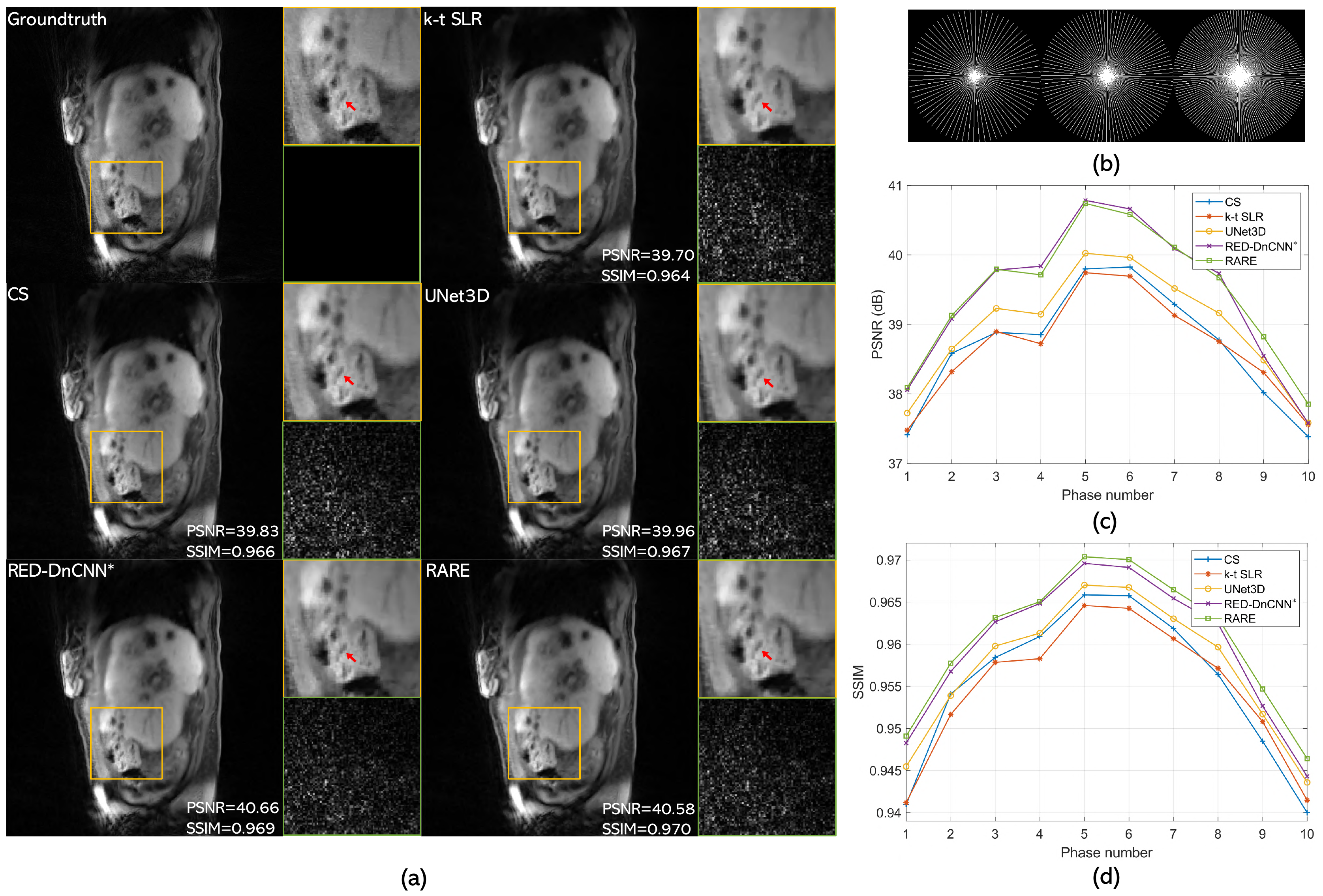}
\end{center}
\caption{Quantitative evaluation of RARE on simulated 4D MRI data: (a) visual comparison of the fifth respiratory phase at 10$\times$ acceleration with 30 dB noise. The yellow box shows an enlarged view of the image. The green box provides the error residual that was amplified $10 \times$ for better visualization. (b) The radial sampling masks used in this simulation, corresponding to 10$\times$, 6.6$\times$, and 5$\times$ accelerations of acquisition. (c) PSNR and (d) SSIM values along respiratory phases for several reference methods. This figure highlights the competitive quantitative performance of RARE compared to several algorithms. Note that the CNN in RARE was trained \emph{without} the groundtruth images used for synthesizing the k-space data.}
\label{Fig:fig1}
\end{figure*}

\begin{figure}[t]
\begin{center}
\includegraphics[width=8.9cm]{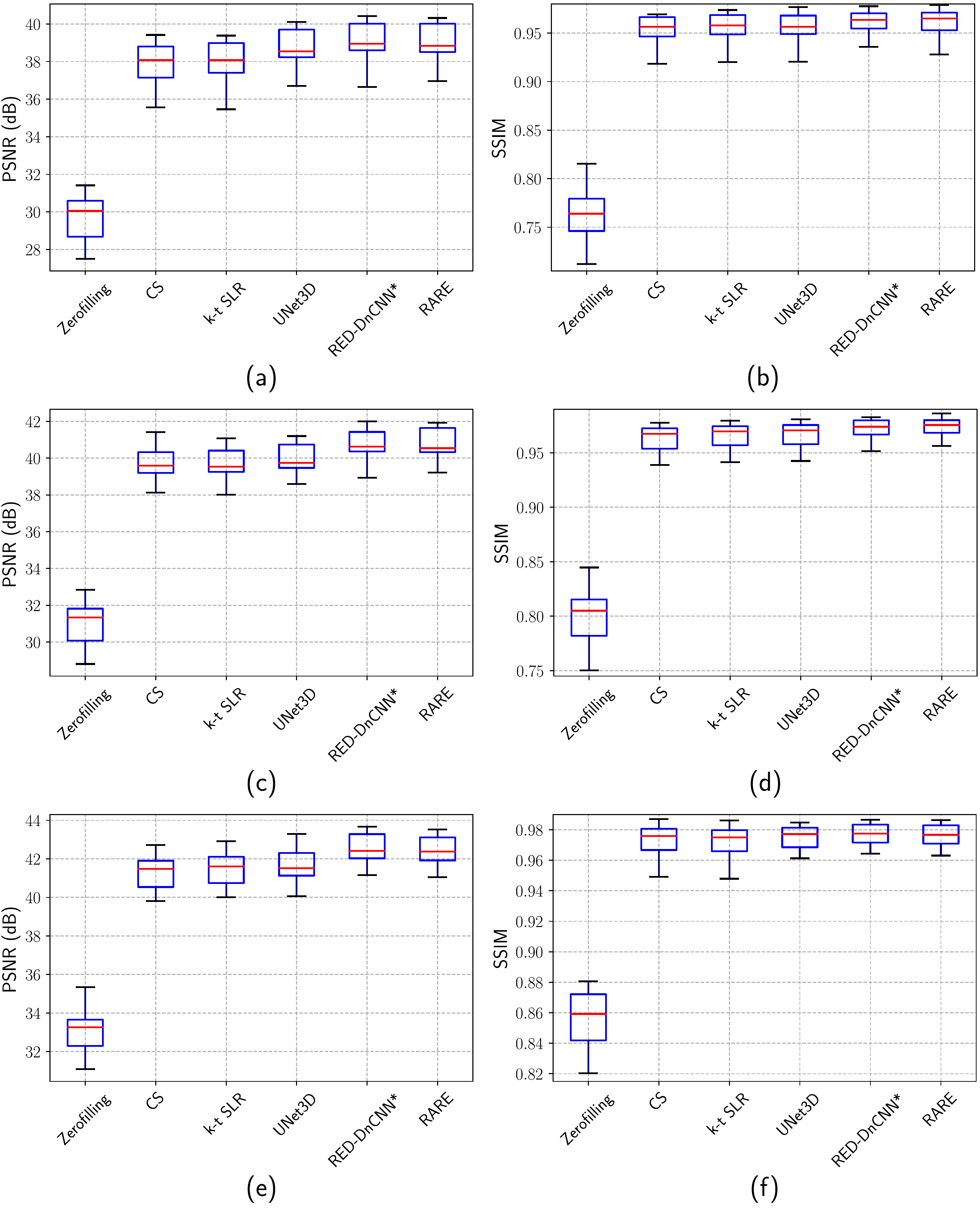}
\end{center}
\caption{The quantitative evaluation of several reference methods on the simulated dataset with noise corresponding to input SNR = 30 dB. From top to bottom, the sampling rates correspond to 10\%, 15\%, and 20\%. Note that unlike the CNNs in UNet3D and RED-DnCNN$^\ast$, the CNN in RARE was trained without the groundtruth images.}
\label{Fig:boxplot}
\end{figure}

\begin{table*}
\centering
\caption{Average PSNR and SSIM values obtained by several reference methods on the simulated dataset. Observe a close agreement between RED and RARE, which highlights the effectiveness of direct training using noisy undersampled data without ground truth. The prior in RED was trained on the groundtruth images used for simulating the measurements, while the one in RARE was trained by using pairs of undersampled and noisy measurements.}
\small
\label{Tab:PSNR}
\scalebox{0.7}{
\begin{tabular*}{22cm}{L{80pt}ccccccccccccccccc} 
\toprule
\textbf{Schemes} & \multicolumn{8}{c}{\textbf{PSNR (dB)}} &  \multicolumn{9}{c}{\textbf{SSIM}}\\
\midrule
\textbf{Methods} & \multicolumn{2}{c}{\textbf{10$\times$}} & & \multicolumn{2}{c}{\textbf{6.6$\times$}} & & \multicolumn{2}{c}{$\textbf{5$\times$}$} & &\multicolumn{2}{c}{\textbf{10$\times$}} & & \multicolumn{2}{c}{\textbf{6.6$\times$}} & & \multicolumn{2}{c}{$\textbf{5$\times$}$} \\
\midrule
& \textbf{30 dB} & \textbf{40 dB} & & \textbf{30 dB} & \textbf{40 dB} & & \textbf{30 dB} & \textbf{40 dB} & &\textbf{30 dB} & \textbf{40 dB} & & \textbf{30 dB} & \textbf{40 dB} & & \textbf{30 dB} & \textbf{40 dB} \\
\cmidrule{2-3} \cmidrule{5-6} \cmidrule{8-9} \cmidrule{11-12} \cmidrule{14-15} \cmidrule{17-18}
\textbf{ZF} 	& 29.54 & 29.67 & & 30.89& 30.98 & & 33.13 & 33.29 & & 0.763 & 0.768 & & 0.801 & 0.805 & & 0.855 & 0.861\\
\textbf{CS} 		& 37.88& 38.22 & & 39.53 & 40.13 & & 41.03 & 42.07 & &0.944 & 0.948 & & 0.955 & 0.961 & & 0.968 & 0.972\\
\textbf{k-t SLR} 	& 37.94 & 38.31 & & 39.49 & 40.02 & & 41.21 & 41.99 & &0.945 & 0.947 & & 0.956 & 0.960& & 0.967 & 0.972\\
\textbf{UNet3D} & 38.46 & 38.80 & & 39.65 & 40.32 & & 41.40 &42.29 & &0.946 & 0.950 & & 0.957 & 0.964 & & 0.969 & 0.974\\
\textbf{RED-DnCNN*} & 38.80& 39.06 & & 40.44 & 40.69 & & 42.17 & 42.79& &0.952 & 0.956& & 0.964 & 0.971& & 0.972&0.976\\
\textbf{RARE} & 38.77 & 39.08 & & 40.46& 40.72 & & 42.08 & 42.65 & & 0.951 & 0.956& & 0.966& 0.970 & & 0.971 & 0.975\\
\bottomrule
\end{tabular*}}
\end{table*}

\begin{figure*}[t]
\begin{center}
\includegraphics[width=16cm]{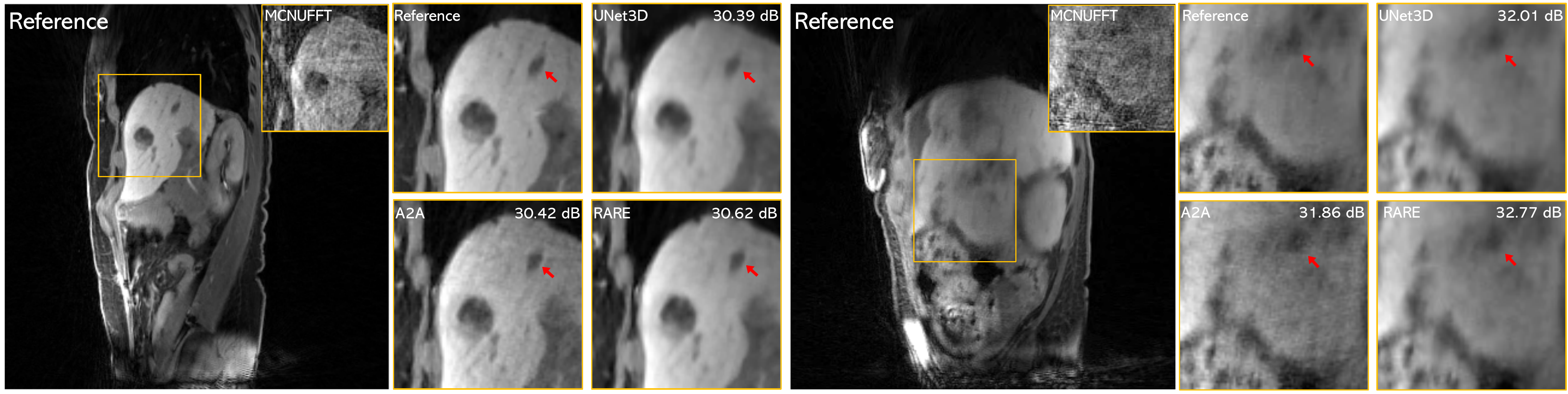}
\end{center}
\caption{Visual illustration of results from experimentally collected data for two patients corresponding to 400 radial spokes (scans of about 1 minute). Several methods, including the artifact-removal CNN (labeled as A2A) and UNet3D are compared against RARE. We also show the results of compressed sensing (CS) reconstruction using 2000-line acquisition (scans of about 5 minutes) as reference. UNet3D was trained to map the 400-line MCNUFFT reconstructions to those by the 2000-line CS. A2A was trained by mapping different MCNUFFT reconstructions to one another. RARE relies on the trained A2A network for regularization. This result shows that RARE can significantly boost the performance of A2A by combining it with the information from the measurement operator. The numbers on the top-right corner correspond to the relative PSNR (rPSNR) values in dB with respect to the reference images.}
\label{Fig:rare_vs_n2n}
\end{figure*}

\begin{figure*}[t]
\begin{center}
\includegraphics[width=16cm]{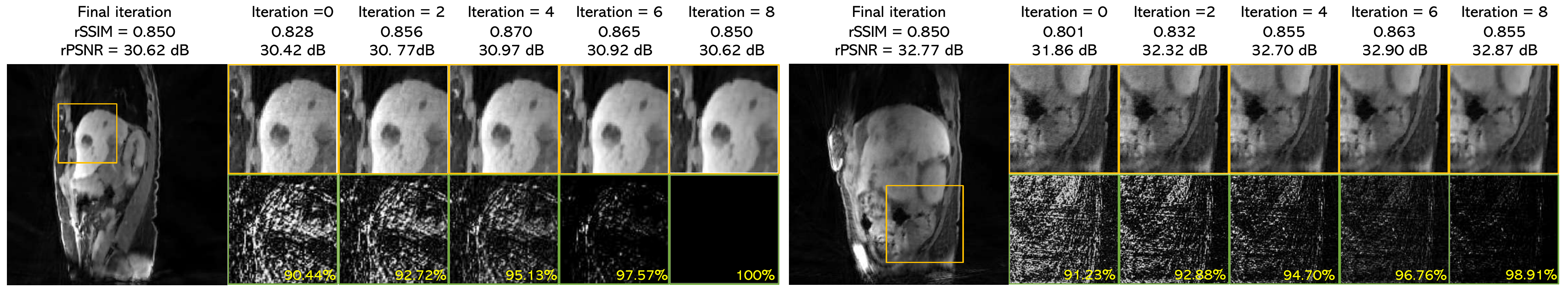}
\end{center}
\caption{Visual illustration of several RARE iterations for two patients starting from the initialization obtained using the artifact-removal CNN (labeled A2A in other figures). The imaging artifact relative to the final solution were magnified $20\times$ for better visualization. The relative change rates are provided inside the residual error and were calculated relative to the final RARE solution. The rPSNR and rSSIM were calculated relative to the reference images shown in Figure~\ref{Fig:rare_vs_n2n}. This figure highlights the potential of RARE to provide fast MR reconstructions (high-quality results achieved within 10 iterations), when initialized with the A2A solution.}
\label{Fig:iter_changes}
\end{figure*}
\begin{figure*}[t]
\begin{center}
\includegraphics[width=16cm]{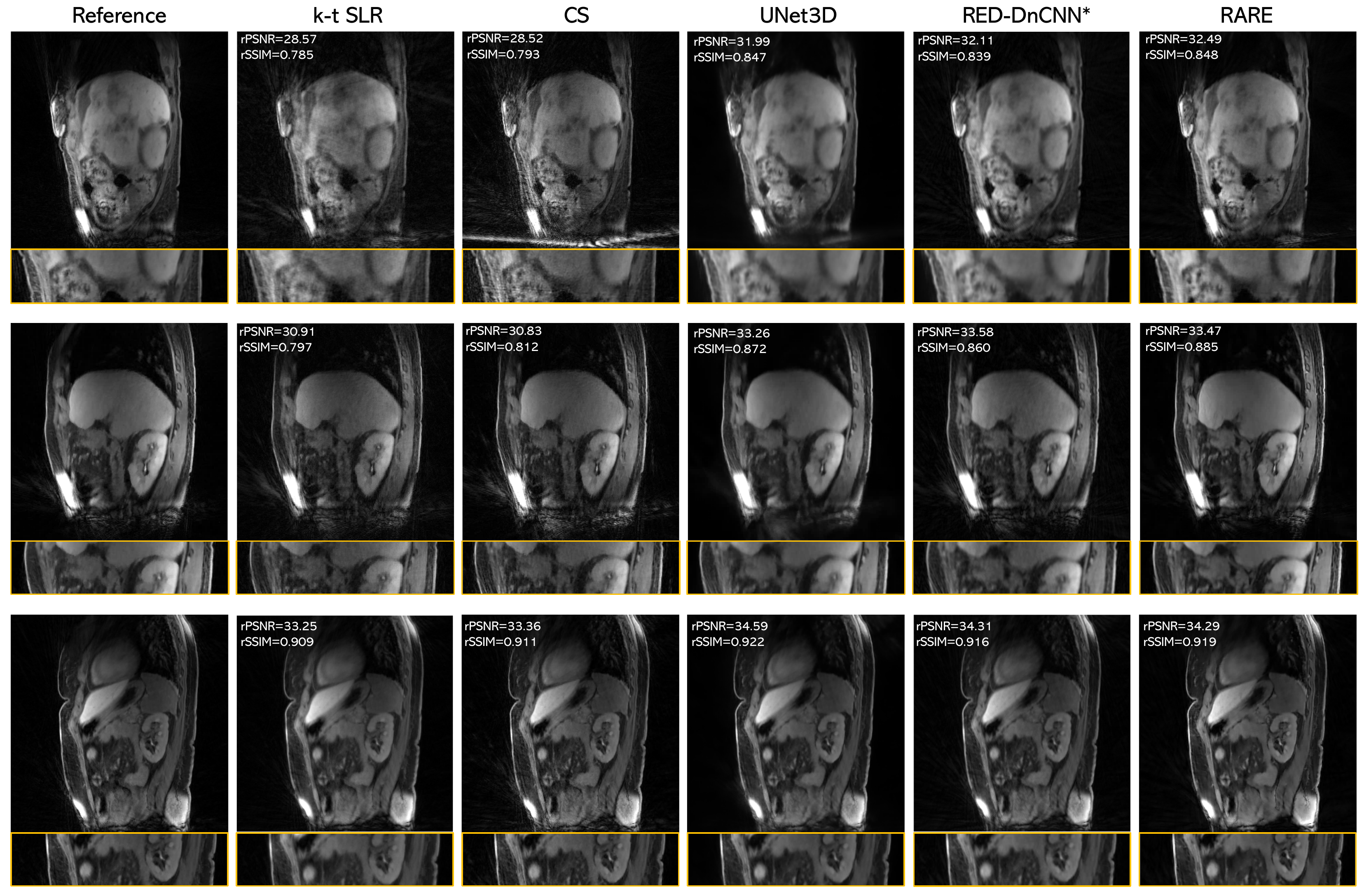}
\end{center}
\caption{Visual illustration of results from experimentally collected measurements on three distinct patients using 400, 800, and 1200 spokes (about 1, 2, and 3 minute scans) from top to bottom, respectively. The leftmost column shows the compressed sensing (CS) reconstruction using 2000 spokes, which is used as reference. The following images correspond to the results of k-t SLR, CS, UNet3D, RED-DnCNN*, and RARE, respectively. This figure highlights the state-of-the-art performance of RARE on 4D MRI reconstruction. Note that since the reference images do not correspond to the groundtruth images, the provided rPSNR and rSSIM values must be interpreted with caution.}
\label{Fig:JSTSP_visual2}
\end{figure*}

\subsection{Implementation for 4D MRI}
The RARE framework for accelerated free-breathing MRI and the architecture for the artifact-removal CNN used in this paper are shown in Figure~\ref{Fig:pipline}.  Since the \emph{Multi-Coil Nonuniform inverse Fast Fourier Transform (MCNUFFT)} data are 4D complex-valued images ($x$, $y$, $z$ for space and $p$ for respiratory phases), we have experimented with different ways of implementing 3D CNNs on 4D data. The best results were obtained when applying the CNN along $(x, y, p)$ dimensions, where $p$ corresponds to different breathing phases within the same $z$-slice. The benefit of this configuration is that the same image features within a $z$-slice move across 10 phases and can be viewed through different artifact patterns.

We used a simplified version of the popular DnCNN denoiser~\cite{Zhang.etal2017} as our artifact-removal prior. The network consists of ten layers with three different blocks. The first block is a composite convolutional layer, consisting of a normal 3D convolutional layer and a rectified linear unit (ReLU) layer, and the second block is a sequence of 8 composite convolutional layers, each having 64 filters. Those composite layers further process the feature maps generated by the first part. The third block of the network, a single convolutional layer, generates the final output image with two channels including the real and imagery part of the image. Every convolution is performed with a stride $=1$, so that the intermediate feature maps have the same spatial size as the input to the CNN. We have also experimented with the UNet3D architecture~\cite{Han.etal2018a, Ronneberger.etal2015} for $\Rsf_\thetabm$. However, we did not observe any quality improvements from UNet3D, making 3D DnCNN a preferable option due to its lower computational complexity compared to UNet3D.

Our implementation of free-breathing 4D MRI is based on the \emph{Consistently Acquired Projections for Tuned and Robust Estimation (CAPTURE)} method~\cite{Eldeniz.etal2018}, which retrospectively bins the k-space measurements into several motion phases. The key observation is that the binning leads to different k-space coverage patterns for different acquisition times, leading to \emph{distinct artifact patterns}. Based on this observation, we learned $\Rsf_\thetabm$ by mapping pairs of complex 3D MR volumes ($x$, $y$ space + $p$ phase) acquired over different acquisition times to one another. Different acquisition times were obtained by splitting our dataset containing 5 minute scans (corresponding to 2000 radial spokes in k-space) of different subjects into the training dataset $\{\xbfhat_{ij}\}$ of 1, 2, 3, 4, and 5 minute MCNUFFT images. Let $(\xbfhat_{ij},\xbfhat_{i'j'})$ in~\eqref{Eq:N2Nlearn}  denote a set of pairs of corrupted images from MCNUFFT reconstructions. The 4D image of the whole body is obtained by applying the model $\Rsf_{\theta}$ slice by slice along $z$. We used a mixed $\ell_1/\ell_2$ loss for training (see eq.~\eqref{Eq:cnn_loss_rare} below). This type of mixture loss function has been shown to be effective in natural image restoration~\cite{Zhao.etal2017}.  Specifically, the training is carried out by minimizing the loss over a training set as follows
\begin{equation}
\begin{aligned}
\label{Eq:cnn_loss_rare}
&\argmin_{\theta}\sum_{i, j, i', j'}\Lcal (\Rsf_{\theta}(\xbfhat_{ij}),\xbfhat_{i'j'}))\quad\text{with}\quad\\\
&\quad\Lcal = \alpha\Lcal_{\ell_1} + (1 - \alpha)\Lcal_{\ell_2},
\end{aligned}
\end{equation}
where $\alpha \in (0,1)$ adjusts the strength of the $\ell_1/\ell_2$ loss.

\section{Numerical Validation}

\label{Sec:Experiments}

We have evaluated RARE qualitatively and quantitatively on simulated as well as experimentally collected \textit{in vivo} liver data at different sampling trajectories. Retrospective downsampling from acquired datasets was used to compare the reconstruction results for each method. All the experiments were performed on a machine equipped with an Intel Xeon Gold 6130 Processor and an NVIDIA GeForce RTX 2080 Ti GPU.

We compared RARE with several widely-used image reconstruction methods, including CS, k-t SLR, UNet3D, and RED-DnCNN$^{\ast}$. The CS implementation~\cite{Eldeniz.etal2018} does 4D space-phase regularization using two complementary regularizers: (a) total variation (TV) that imposes piecewise smoothness across different respiratory phases and (b) total generalized variation (TGV) regularization that imposes higher-order piecewise smoothness in the spatial domain. Both regularizers are weighted by the regularization parameters that were optimized for the performance. The k-t SLR algorithm\footnote{The code for k-t SLR is publicly available at \url{http://user.engineering.uiowa.edu/~jcb/software.html}.} exploits the global low-rank structure of the spatiotemporal signal~\cite{lingalaetal.2011}. Specifically, the problem is posed as a matrix recovery problem using a low-rank penalty with an additional sparsity-promoting prior to improve the reconstruction quality. UNet3D corresponds to our own implementation of the architecture used in~\cite{Sun.etal2018}. The network was trained in the usual supervised fashion~\cite{DJin.etal2017} using the $\ell_2$ loss. Additional details on training for synthetic and in-vivo experiments are provided in dedicated sections below. The network has an analysis and a synthesis path, each with four resolution steps. In the analysis path, each layer contains three 3\,$\times$\,3\,$\times$\,3 convolutions each followed by a ReLU, and then a 1\,$\times$\,2\,$\times$\,2 max pooling with strides of two in $x$-$y$ axises while keeping the phase dimension unchanged. In the synthesis path, each layer consists of an up-convolution of 2\,$\times$\,2\,$\times$\,2 by strides of two in each dimension, followed by three 3\,$\times$\,3\,$\times$\,3 convolutions each followed by a ReLU. We adopted the DnCNN$^\ast$ architecture proposed in~\cite{Ahmad.etal2019} as the AWGN denoiser for RED. The network has seven layers, including 5 hidden layers, an input layer, and an output layer. We have experimented with higher number of DnCNN$^\ast$ layers, but observed that additional layers beyond 7 do not improve the imaging quality when the CNN is used within RED.

Two widely used quantitative metrics were implemented for comparing different algorithms: the \emph{peak signal-to-noise ratio (PSNR)} in dB and the \emph{structural similarity index (SSIM)}
\begin{equation}
\label{Eq:SSIM}
\hbox{SSIM}(\xbm,\ybm) = \frac{(2\mu_x\mu_y + c_1)(2\sigma_{xy} + c_2)}{(\mu_x^2 + \mu_y^2 + c_1)(\sigma_x^2 + \sigma_y^2 + c_2)},
\end{equation}
where $\mu$ and $\sigma$ are the average and variance of the image. $c_1 = (k_1L)^2$, $c_2 = (k_2L)^2$ are two variables to stabilize the division with weak denominator and $L =\,$dynamic range of the pixel-values. For the results on simulated data, we compute the PSNR and SSIM values with respect to the groundtruth images used to synthesize the measurements. Since there is no fully-sampled groundtruth in experimental data, we quantify the performance with respect to the reference images $\xbm_{\text{ref}}$ that were obtained using the 2000-line CS reconstruction. We label these values \emph{relative SSIM (rSSIM)} and \emph{relative PSNR (rPSNR)} to highlight that they must be interpreted with caution.

\begin{figure*}[t!]
\begin{center}
\includegraphics[width=16cm]{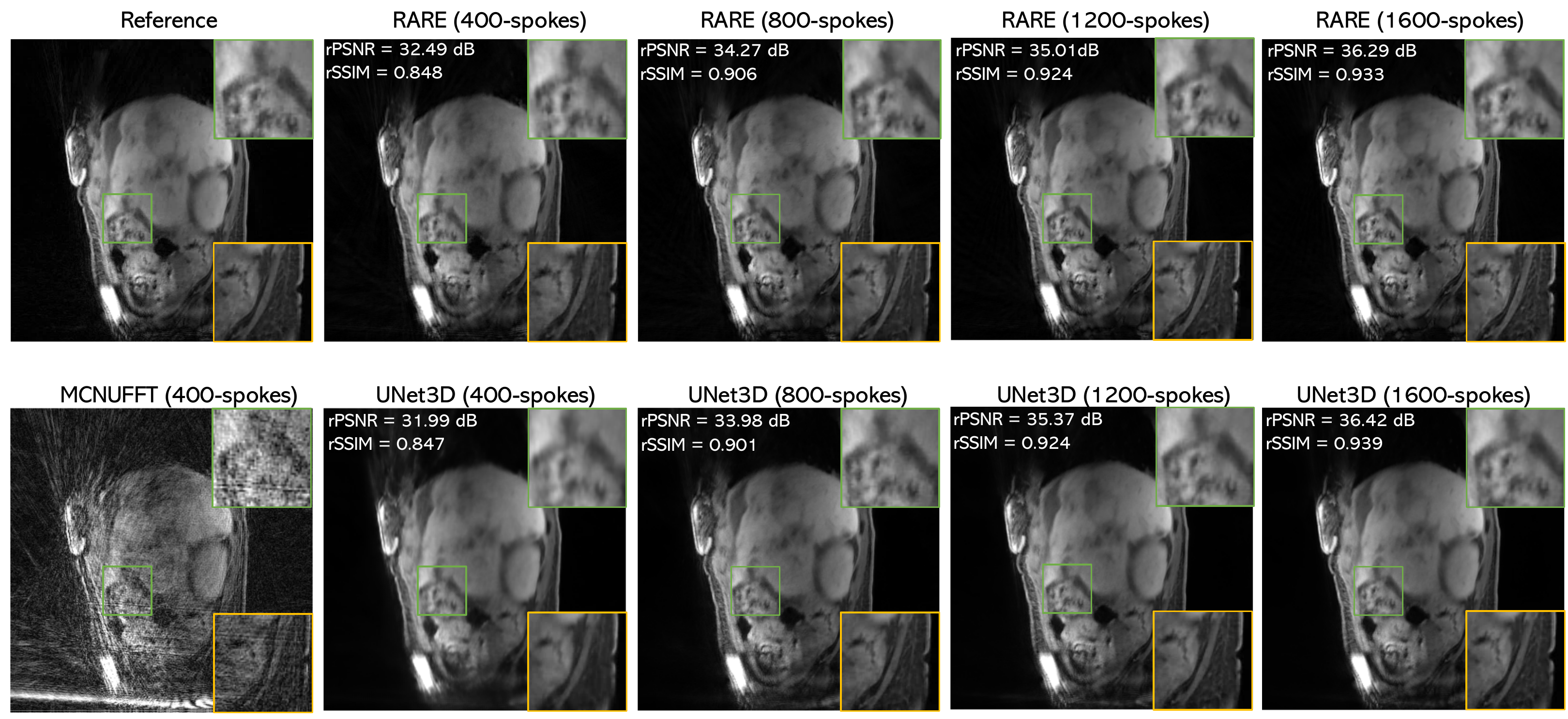}
\end{center}
\caption{The effect of additional data on the performance of RARE and UNet3D. UNet3D is trained to map 400-line MCNUFFT images to the reference images obtained using 2000-line CS reconstruction (leftmost column). On the other hand, the prior in RARE was trained only by using MCNUFFT data. This figure highlights the ability of RARE to achieve high-quality results under severe undersampling rates. Note that since the reference images do not correspond to the groundtruth, the rPSNR and rSSIM values must be interpreted with caution.}
\label{Fig:line_flow}
\end{figure*}

\subsection{Results on Simulated Data}

We have quantitatively validated RARE against several baseline methods by using a dataset synthesized on simulated radial measurements. The groundtruth corresponds to the magnitude images obtained from CAPTURE-CS on 2000 radial spokes. The image from the first 8 healthy subjects were used for training, 1 for validation, and the rest were used for testing. The complete 4D image volume in simulations consists of 14 $z$-slices with each slice containing a 3D image volume of size 10\,$\times$\,320\,$\times$\,320.

We performed retrospective radial undersampling in order to simulate a practical single-coil acquisition scenario. The forward model is $\ybm = \Hbm\xbm + \ebm,$ where ${\ebm}$ is an AWGN vector and ${\Hbm}$ is a matrix corresponding to the radially-subsampled Fourier transform. We set the sampling rate to 10\%, 15\%, and 20\% of the original k-space data (corresponding to 10$\times$, 6.6$\times$, 5$\times$  acceleration) with AWGN corresponding to input SNR of 30 dB and 40 dB. To quantitively compare each methods, the PSNR and SSIM values were calculated with respect to the groundtruth images used to synthesize the measurements. In all the experiments below, RARE was run with the stopping parameter $\rho = 10^{-6}$ and the maximal number of $500$ iterations.




In this experiment, we trained the artifact-removal network using pairs of two independent undersampled images corresponding to the sampling rate of 40\% with AWGN corresponding to 30 dB and 40 dB using the $\ell_2$ loss. For the CNN denoiser, it was trained for the removal of AWGN at four noise levels corresponding to $\sigma\in$ \{1, 3, 5, 10\}. For each experiment, we selected the denoiser achieving the highest PSNR value. Several instances of UNet3D were trained for each sampling rate and input SNR levels. For k-t SLR and CS, we performed grid search to identify the optimal parameters for the data. Since the synthetic groundtruth images were real magnitude images, we only kept the real part of the gradient step during each iteration for both RARE and RED-DnCNN$^{\ast}$.

Figure~\ref{Fig:boxplot} and Table~\ref{Tab:PSNR} summarize the quantitative results in terms of PSNR and SSIM. Overall, all methods achieved good performance, with learning-based methods (UNet3D, RED, and RARE) achieving the best performance. Moreover, RARE achieves comparable imaging quality to RED-DnCNN$^{\ast}$, which is remarkable considering the fact that RARE doesn't use the groundtruth in training. As a remider, the prior in RARE is trained only on zero-filled reconstructions, while the one in RED on the actual groundtruth images used for synthesizing the measurements. Figure~\ref{Fig:fig1} shows some representative visual results for the synthetic liver test data, using various reconstruction methods at 10$\times$ acceleration with input SNR = 30 dB. The error plots were magnified by 10$\times$ and highlighted with gray color map. In general, the reconstructions of learning based methods outperform the k-t SLR and CS. We observe that k-t SLR produces cartoon-like features in the image. Although CS shows high denoising quality, many details are not preserved compared with learning based methods.

\subsection{Results on Experimental Data}

The data acquisition was performed on a 3T PET/MRI scanner (Biograph mMR; Siemens Healthcare, Erlangen, Germany). The measurements were collected using the CAPTURE method~\cite{Eldeniz.etal2018}, a recently proposed T1-weighted stack-of-stars 3D spoiled gradient-echo sequence with fat suppression that has consistently acquired projections for respiratory motion detection. Upon the approval of our Institutional Review Board, multi-channel liver data from 10 healthy volunteers and 6 cancer patients were used in this paper.
The acquisition parameters were as follows: TE/TR = 1.69 ms/3.54 ms, FOV = 360 mm\,$\times$\,360 mm, in-plane resolution = 1.125\,$\times$\,1.125 mm, slice partial Fourier factor = 6/8, number of radial lines = 2000, slice resolution = 50\%, slices per slab = 96, 112 or 120 so as to cover the torso with a slice thickness of 3 mm, and total acquisition time = about 5 minutes (slightly longer for larger subjects). We  drop  the  first  10  spokes during   reconstruction  in  order  to make sure the acquired signal reaches a steady state. Our free-breathing MRI data was then subsequently binned into 10 respiratory phases, and thus each phase view was reconstructed with 199 spokes accordingly. The dimensions of each $z$-slice image are 320\,$\times$\,320\,$\times$\,10 for each subject. The coil sensitivity maps were estimated from the central radial k-space spokes of each slice and were assumed to be known during experiments. Apodization was applied by using a Hamming window that covers this range in order to avoid Gibbs ringing. We then implemented MCNUFFT on those individual coil element data. 

We used the first 8 healthy subjects for training and 1 for validation. The complex-valued 4 dimensional images were cropped along $x$-$y$ axes to obtain images of parts of the anatomy. The real and imaginary parts of these images were separated into two input channels of the artifact-removal CNNs. Thus, our training and validation data were structured as slices\,$\times$\,phases\,$\times$\,rows\,$\times$\,columns\,$\times$\,channels as $N_s \,\times$\,10\,$\times$\,320\,$\times$\,320\,$\times$\,2 for each subject, where $N_s$ is 96, 112, or 120 depending on the subject. For testing, we picked 5 slices of the remaining healthy subject and of each patient. We conducted the experiments for various acquisition durations 1, 2, 3, 4, and 5 minutes, corresponding to 400, 800, 1200, 1600, and 2000 radial spokes in k-space, respectively. We note that the multi-coil sensitivity maps for each downsampling rate was recalculated accordingly. 

We used images reconstructed using CS from 2000 radial spokes (corresponding to 5 minute scans) as our \emph{reference} images for quantitative evaluations and training of baseline CNNs~\cite{Eldeniz.etal2018}. The quality of reconstructed images for all algorithms were quantified with respect to reference images by using rSSIM and rPSNR in dB. The regularization parameters of iterative algorithms in the following experiments were optimized quantitatively to maximize rPSNR. In all the experiments below, RARE was run with the stopping parameter $\rho = 10^{-2}$ and the maximal number of $30$ iterations.

Figure~\ref{Fig:rare_vs_n2n} illustrates reconstructions for 400 radial spokes (corresponding to 1 minute acquisition) for two patients. We compare RARE with the results of direct application of artifact-removal CNNs, referred to \emph{Artifact2Artifact} (A2A) in the figure. The comparison area was highlighted with a yellow box for each patient. The MCNUFFT reconstruction of the highlighted area is also shown, along with the reference images corresponding to CAPTURE-CS reconstructions from 2000 spokes. UNet3D was trained by mapping 400 spoke MCNUFFT reconstructions to the reference images. RARE was initialized with the result of A2A. One can observe that all methods yield significant improvements over the MCNUFFT reconstruction. A closer inspection shows that UNet3D leads to image blurring, which reduces the contrast of the tumor highlighted by the red arrow. On the other hand, A2A leads to streaking artifact that we attribute to the fact that it was trained directly on undersampled and noisy images obtained via MCNUFFT. RARE significantly boosts the performance of A2A, by leveraging information from the measurement operator, which leads to sharper results. We also tested CS, k-t SLR, and RED-DnCNN$^{\ast}$ methods on the same data, but omitted them from the figure, choosing UNet3D and A2A as the representative methods for this figure (see Figure~\ref{Fig:JSTSP_visual2} for the performance of other methods).

Figure~\ref{Fig:iter_changes} demonstrates the evolution of the iterates generated by RARE over 8 iterations after initialization by A2A. The final result of RARE shown here is identical to the one in Figure~\ref{Fig:rare_vs_n2n}. The yellow box provides an enlarged view from the image, while the green box shows the residual with respect to the final iteration. The difference images were amplified $20\times$ for better visualization of artifact patterns. We observed  no significant visual improvements after 10 iterations of RARE. The figure highlights that RARE is effective in reducing the streaking artifacts apparent in the A2A solution.

Figures~\ref{Fig:JSTSP_visual2} and~\ref{Fig:line_flow} show several visual results comparing the performance of RARE on different patients and acquisition lengths, against several baseline methods, including CS, k-t SLR, UNet3D, and RED-DnCNN$^{\ast}$. We conducted experiments for 400, 800, 1200, and 1600 radial spokes, corresponding to roughly 1, 2, 3, and 4 minute acquisitions, respectively.   The solution of A2A was used as the initialization for RARE. For RED-DnCNN$^{\ast}$, we trained the CNN denoiser for the removal of complex-valued AWGN with $\sigma\in \{0.5, 1, 1.5, 2\}$. Parameters used in different methods were optimized for the maximum rPSNR value. Note that all methods achieve similar visual results for 1600 spokes, and thus we choose 400, 800, and 1200 spokes reconstruction as the representative examples in the figure. For 400 and 800 spokes reconstruction, k-t SLR and CS still show streaking artifacts. Although UNet3D and RED-DnCNN$^{\ast}$ are able to remove these artifact, RARE provides improved artifact removal while also producing sharper images with a better contrast. For 1200 spokes, all the algorithms show significant improvements, but RARE shows slightly better reconstruction results compared with other learning based methods in terms of the visual quality. Figure~\ref{Fig:line_flow} highlights a noticeable improvement in quality for 800 spokes for both RARE and UNet3D, when compared to 400 spokes. Although, UNet3D yields higher rPSNR on 800 spokes with respect to the \emph{reference} images (which do not correspond to groundtruth), RARE still provides high quality visual results with more details, highlighted in the zoomed regions. Note that the distinct performance of RED and RARE is not surprising on experimental data. While both methods have similar algorithmic structure, they rely on distinct priors that were trained using fundamentally different learning strategies. The prior in RED was trained by using the reference images as labels, while the one in RARE uses the A2A network. Our experimental results highlight the potential of training the imaging prior directly on the MCNUFFT reconstructed images, thus bypassing the need to generate training labels by using the computationally expensive CS reconstruction.


\section{Conclusion}
\label{Sec:Conclusion}
We proposed RARE as a method for image reconstruction that uses priors learned directly from undersampled and noisy data. We provided extensive experimental results for motivating practical relevance of RARE. Our results indicate that RARE provides competitive image quality compared to several baseline methods. While our experiments focused on MRI, the method is broadly applicable to many other imaging modalities, where it is easy to collect several distinct views of an object for training.  The method presented in this paper can be extended in several complementary ways. An interesting direction would be to consider the inclusion of manifold or low-rank priors for exploiting non-local redundancies in the time dimension, as was done in~\cite{Biswas.etal2019}. Another interesting direction would be to consider applications of RARE to other imaging problems, such as computerized tomography (CT)~\cite{DJin.etal2017} and optical diffraction tomography (ODT)~\cite{Kamilov.etal2016}, where it is often impossible to obtain fully-sampled measurements.

\bibliographystyle{IEEEtran}


\end{document}